\documentclass[fleqn,usenatbib]{mnras}

\usepackage[T1]{fontenc}

\DeclareRobustCommand{\VAN}[3]{#2}
\let\VANthebibliography\thebibliography
\def\thebibliography{\DeclareRobustCommand{\VAN}[3]{##3}\VANthebibliography}

\usepackage{graphicx}	% Including figure files
\usepackage{amsmath}	% Advanced maths commands
\usepackage{amssymb}	% Extra maths symbols
\usepackage{caption}
\usepackage{subcaption}
\usepackage{pdflscape}
\usepackage{blindtext, multicol}
\usepackage[export]{adjustbox}
\newcommand{\dmunits}{\,pc\,cm$^{-3}$}
\newcommand{\filtool}{\mbox{\texttt{filtool}}}
\newcommand{\candidatefilter}{\mbox{\textsc{candidate\_filter}}}
\newcommand{\peasoup}{\mbox{\textsc{peasoup}}}

\newcommand{\pulsarx}{\mbox{\textsc{pulsarX}}}
\newcommand{\mosaic}{\mbox{\textsc{mosaic}}}
\newcommand{\dedisp}{\mbox{\textsc{dedisp}}}
\newcommand{\transientx}{\mbox{\textsc{transientX}}}
\newcommand{\candyjar}{\mbox{\textsc{CandyJar}}}
\newcommand{\presto}{\mbox{\textsc{presto}}}
\usepackage{newtxtext,newtxmath}
\title[Transient search in NGC~253 with MeerKAT]{Searching for Pulsars, Magnetars, and Fast Radio Bursts in the Sculptor Galaxy using MeerKAT}

\author[H. Hurter et al.]{
H. Hurter,$^{1}$\thanks{E-mail: heinrich.hurter11@gmail.com}
C. Venter,$^{1,2}$
L. Levin,$^{3}$
B. W. Stappers$^{3}$
E. D. Barr,$^{4}$
R. P. Breton,$^{3}$
S. Buchner,$^{5}$
E. Carli,$^{3}$ \newauthor
M. Kramer,$^{4}$
P.~V.~Padmanabh$^{6,7,4}$
A. Possenti,$^{8}$
V. Prayag,$^{9,10}$
J. D. Turner$^{3}$
\\ \\ \\
$^{1}$ Centre for Space Research, North-West University, Private Bag X6001, Potchefstroom 2520, South Africa\\
$^{2}$ National Institute for Theoretical and Computational Sciences (NITheCS), Stellenbosch, South Africa \\
$^{3}$ Jodrell Bank Centre for Astrophysics, Department of Physics and Astronomy, The University of Manchester, Manchester, M13 9PL, UK\\
$^{4}$ Max-Planck-Institut f\"ur Radioastronomie, Auf dem H\"ugel 69, D-53121 Bonn, Germany\\
$^{5}$ South African Radio Astronomy Observatory (SARAO), 2 Fir Street, Black River Park, Observatory, Cape Town 7925, South Africa\\
$^{6}$ Max Planck Institute for Gravitational Physics (Albert Einstein Institute), D-30167 Hannover, Germany \\
$^{7}$ Leibniz Universit\"{a}t Hannover, D-30167 Hannover, Germany \\
$^{8}$ INAF-Osservatorio Astronomico di Cagliari, via della Scienza 5, \=I-09047 Selargius (CA), Italy \\
$^{9}$ Department of Astronomy, The University of Cape Town, Private Bag X3, Rondebosch 7701, Cape Town, South Africa\\
$^{10}$ High Energy Physics, Cosmology \& Astrophysics Theory (HEPCAT) Group, Department of Mathematics \& Applied Mathematics, \\University of Cape Town, Cape Town 7700, South Africa
}

\date{Accepted XXX. Received YYY; in original form ZZZ}

\pubyear{2024}

\begin{document}
\label{firstpage}
\pagerange{\pageref{firstpage}--\pageref{lastpage}}
\maketitle

\begin{abstract} 
The Sculptor Galaxy (NGC~253), located in the Southern Hemisphere, far off the Galactic Plane, has a relatively high star-formation rate of about 7~M$_{\odot}$ yr$^{-1}$ and hosts a young and bright stellar population, including several super star clusters and supernova remnants. It is also the first galaxy, apart from the Milky Way Galaxy to be associated with two giant magnetar flares. As such, it is a potential host of pulsars and/or fast radio bursts in the nearby Universe. The instantaneous sensitivity and multibeam sky coverage offered by MeerKAT therefore make it a favourable target. We searched for pulsars, radio-emitting magnetars, and fast radio bursts in NGC~253 as part of the TRAPUM large survey project with MeerKAT. We did not find any pulsars during a four-hour observation, and derive a flux density limit of 4.4~$\upmu$Jy at 1400~MHz, limiting the pseudo-luminosity of the brightest putative pulsar in this galaxy to 54~Jy kpc$^2$. Assuming universality of pulsar populations between galaxies, we estimate that detecting a pulsar as bright as this limit requires NGC~253 to contain a pulsar population of $\sog$20 000. We also did not detect any single pulses and our single pulse search flux density limit is 62~mJy at 1284\,MHz. Our search is sensitive enough to have detected any fast radio bursts and radio emission similar to the brighter pulses seen from the magnetar SGR~J1935+2154 if they had occurred during our observation.
\end{abstract}
%%%%%%%%%%%%%%%%%%%%%%%%%%%%%%%%%%%%%%%%%%%%%%%%%%

%%%%%%%%%%%%%%%%% BODY OF PAPER %%%%%%%%%%%%%%%%%%
\section{Introduction} \label{sec:introduction}
%%%%%%%%%%%%%%%%%%%%%%%%%%%%%%%%%%%%%%%%%%%%%%%%%%%

The detection and monitoring of radio pulsars, given their powerful magnetic fields, extreme compactness, and broadband electromagnetic radiation, are important in a variety of disciplines such as exploring General Relativistic effects \citep{1979Natur.277..437T}, the equation of state of neutron stars (NSs; e.g., \citealt{Koehn2024}), as well as relativistic plasma and radiation physics \citep[e.g.,][]{Philippov2022}.
Currently, the Milky Way (MW) galaxy hosts more than 99 per cent (3\,608 to date) of the known radio pulsar population \citep{manchester2005australia}\footnote{ATNF catalogue version 2.3.0: \url{https://www.atnf.csiro.au/research/pulsar/psrcat/}}. The remaining 41 pulsars are extragalactic and were discovered in two satellite galaxies of the MW, the Magellanic clouds (e.g. \citealt{2006ApJ...649..235M,2001ApJ...553..367C,2022ApJ...928..161H,2019MNRAS.487.4332T,2024MNRAS.tmp.1306C}). The Magellanic clouds are considered to be good targets for pulsar searches, since they are not obstructed by the Galactic plane, and therefore searches are not greatly affected by dispersive effects of the Galactic interstellar medium. 
Furthermore, the Magellanic Clouds host diverse NS populations that are indicative of various evolutionary stages of massive stars, and a large number of supernova remnants (see e.g., \citealt{2020MNRAS.494..500T,Badenes_2010}), which further motivates the pulsar searches in these types of environment. NSs are created by core-collapse events of stars with masses between $8-25 {\rm M}_{\odot}$ depending on their metallicity and binary companion \citep{2003ApJ...591..288H}.
Considering factors such as star formation history, stellar mass distributions, and stellar types of galaxy targets is important for pulsar searches. For instance, recent star-forming episodes produce populations of NSs that may be detectable as radio pulsars. 

An increasing number of fast radio bursts (FRBs) have been detected in the last few decades. FRBs are classified by their one-off or multiple short-duration bursts \citep{2007Sci...318..777L, 2013Sci...341...53T, 2016Natur.531..202S}, and it is accepted that FRBs are extragalactic events due to the high associated dispersion measure (DM) that put them well beyond the MW limits (e.g., \citealt{2019PhR...821....1P}). The mechanisms responsible for repeating and non-repeating bursts are still a matter of debate. With the detection and localisation of more bursts, constraints can be placed on their emission mechanism (e.g., \citealt{2021ApJS..257...59C}). A possible explanation for non-repeating FRBs is that they are an after-effect of cataclysmic events (e.g., \citealt{2020A&A...639A.119M}). On the contrary, repeating FRBs should have a non-cataclysmic origin \citep{2016Natur.531..202S}. A recent detection of FRB-like bursts was made from the Soft-Gamma Repeater (SGR)~J1935+2154, which is a Galactic magnetar. This suggests a strong connection between magnetar emission and at least some repeating FRBs \citep{2020Natur.587...59B,2020Natur.587...54C,2006Natur.442..892C}. Furthermore, SGRs are associated with regions of recent star-forming and supernovae \citep{GAENSLER2004645}.
Known FRB host galaxies exhibit a wide range of compositions, with masses in the range ${\rm M}_{*}=10^8$ -- $10^{10} \,{\rm M}_{\odot}$ and star-formation rates (SFRs) ranging from $0.05$ -- $10 \,{\rm M}_{\odot}{\rm yr}^{-1}$ \citep{bhandari2022characterizing}. \citet{mannings2021high} found that all FRBs located in hosts with spiral structures occur near or on a spiral arm.

The Nearby Galaxies programme within the TRansients And PUlsars with MeerKAT (TRAPUM\footnote{\url{https://www.trapum.org}}) Large Survey Project \citep{2016mks..confE...9S}, has as one of its aims the discovery of new pulsars and fast transients outside of the MW galaxy, and to investigate their properties in relation to those of their host galaxy. One of the targets of this survey is NGC~253, also known as the Sculptor Galaxy. It is located in the southern sky far below the Galactic Plane, at a Galactic latitude of $b= -87.96^\circ$ and a distance of $3.5\pm0.2$~Mpc \citep{rekola2005distance}. It is a promising target for finding pulsars, due to its population of massive stars and its similarity to the MW. 
It is classified as a barred spiral galaxy and is one of the nearest galaxies undergoing a nuclear starburst. It is one of the brightest starburst galaxies in the Southern Hemisphere, with detections of numerous supernova remnants (SNRs) within the starburst region, including one optically identified supernova (SNR 1940E, \citealt{2018MNRAS.474.4937C,Zwicky}) close to the central region of the galaxy. The unobscured stellar population is consistent with ages  $<8$~Myr \citep{2009ApJ...697.1180K,2016ApJ...818..142D}. The centre hosts a population of more than a dozen super star clusters (SSC) that are still in the formative stages. A near-infrared photometry study of this galaxy revealed 181 star clusters in the central 600 pc \citep{2022BAAA...63..232C}. Nine of these clusters have masses in the range of 10$^5 - 10^6$ M$_{\odot}$ and ages less than 7~Myr, with the remaining clusters being older and less massive. This is interesting since these clusters can provide insights into the star formation history of this galaxy. NGC~253 has a total mass of $(8.1\pm2.6)\times 10^{11}$ $ {\rm M}_{\odot}$ \citep{2021AJ....161..205K}, which is about a factor of two less than the total mass of the MW. 
\citet{1993ApJ...406L..11F} suggested that the central 6$\arcsec$ of NGC~253 contains about 24,000 O-type stars and that SNRs are located throughout the nuclear region. The nucleus has an SFR of approximately $5 \,{\rm M}_{\odot}{\rm yr}^{-1}$, which accounts for about 70 per cent of the rate of the entire galaxy \citep{wik2014spatially}. 
The central SFR of NGC~253 is 30--40 times higher than that of the central region of the MW \citep{2013MNRAS.429..987L,2017MNRAS.469.2263B,2021ApJ...919..105M}. 
A high SFR is consistent with a young stellar population \citep{2022A&A...666A.186D}, which in turn is likely to contribute to a large SNR population. Furthermore, \citet{wik2014spatially} found that X-ray emission above $\sog$10~keV were concentrated in the central 100$\arcsec$ of NGC 253, produced by three nuclear sources, an off-nuclear ultraluminous X-ray source and a pulsar candidate, which could be an extremely luminous X-ray pulsar.
NGC~253 is furthermore the first galaxy to be associated with two magnetar giant flares (MGFs), (Gamma-Ray Burst) GRB~180128A and GRB~200415A (\citealt{2024A&A...687A.173T,2021Natur.589..211S}). GRB~200415A was observed on April 15, 2020, located in an area of 20 arcmin$^2$ that overlaps the central region of NGC~253 \citep{2021Natur.589..211S}. GRB~180128A occured 808 days before GRB~200415A on January 28, 2018, and was found in archival data after the discovery of GRB~200415A. However, it is still unclear whether these bursts originated from the same magnetar \citep{2024A&A...687A.173T}. The formation channel of magnetars is consistent with core-collapse supernovae \citep{heintz2020host,bochenek2021localized}. Thus, the high SFR, the presence of SNRs, the high energy X-ray sources, young star clusters, and associated MGFs support the likelihood that NGC~253 contains a young NS/pulsar population. It is also interesting to note that young pulsars sometimes emit giant radio pulses \citep{1995ApJ...453..433L,2021MNRAS.505.4468G}, which would enable us to discover distant radio pulsars that might otherwise not be detectable \citep{2003ApJ...596..982M}. The probable presence of at least one magnetar in this galaxy also means that there may be sources responsible for emitting FRB-like pulses, like those seen from SGR~J1935-2154.

Extragalactic targets such as NGC~253, NGC~300, NGC~6300, NGC~7793, and Fornax were previously searched for giant radio pulses using a single beam from the Parkes (Murriyang) Telescope, but none were detected \citep{2003ApJ...596..982M}. Two pulsar search surveys (see \autoref{sec:discussion_and_conclusion}) also conducted short integration pulsar searches in the direction of NGC~253. However, these searches did not yield evidence for pulsars in NGC~253. 

We made use of the exceptional sensitivity of the MeerKAT Radio Telescope to perform a new deeper pulsar and FRB search. The MeerKAT Radio Telescope is a radio interferometer located in the Northern Cape region of South Africa and has remarkable sensitivity for pulsar and fast transient searches. MeerKAT is made up of 64 parabolic antennas, each with a diameter of 13.5~m, distributed to form a maximum baseline of 8~km \citep{2016mks..confE...1J,2018NatAs...2..594C}. 
The gain of the fully phased array is about 4 times higher than that of the Parkes Telescope and 1.4 times higher than that of the Green Bank Telescope \citep{bailes2020meerkat}, therefore making the MeerKAT Radio Telescope even more suitable for extragalactic transient searches.

In this article we describe our observation of NGC~253 and data reduction methods in \autoref{sec:observation}. We present our results and search sensitivities in \autoref{sec:resutls}. Our conclusions follow in \autoref{sec:discussion_and_conclusion}.

\begin{figure}
    \centering
    \includegraphics[scale=0.2,width=0.5\textwidth]{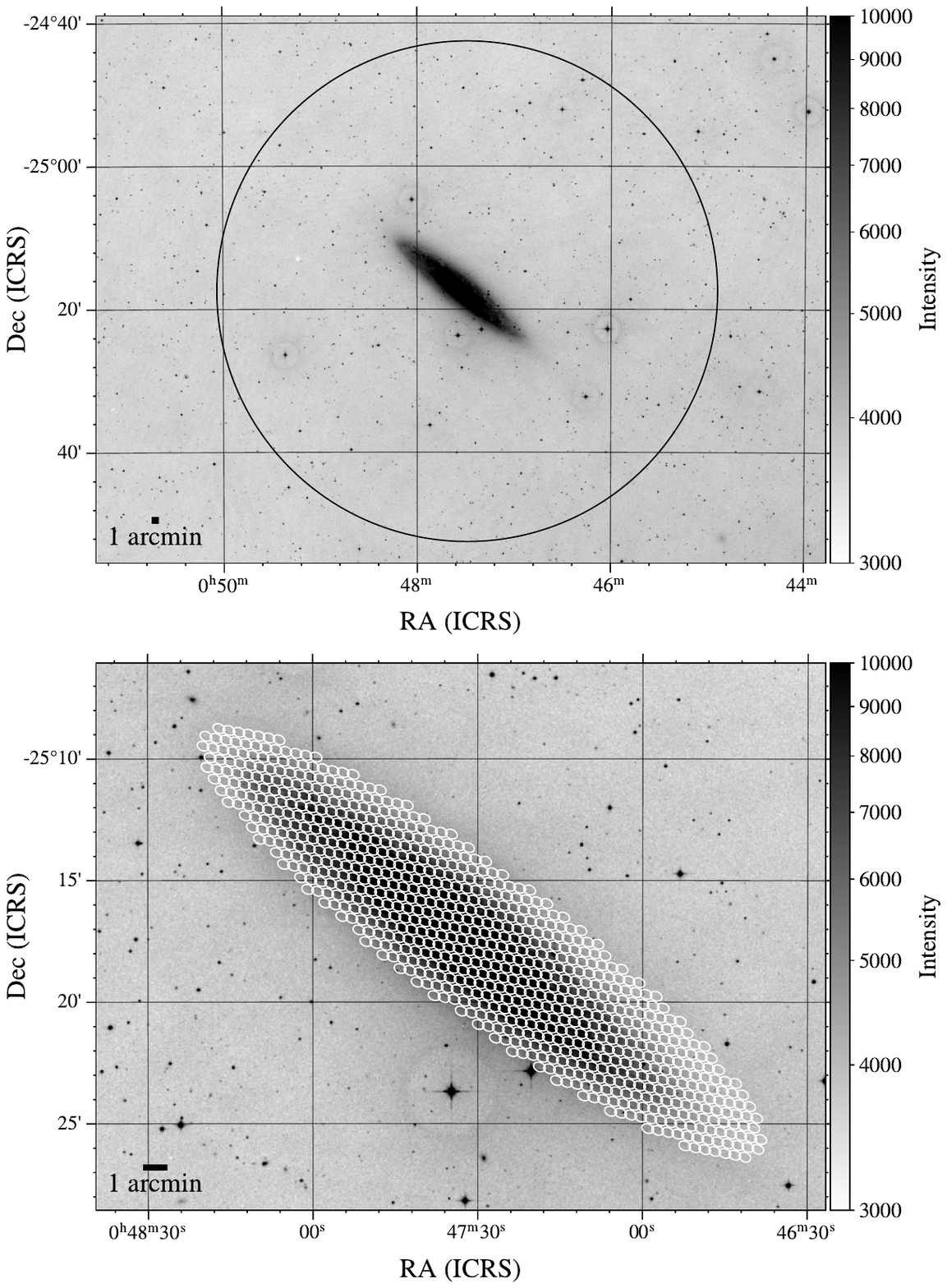}
    \caption{An image of NGC~253 from the Digitised Sky Survey (DSS2) in the near-infrared (NIR) band. The incoherent beam can be seen in the upper plot, and was centered on SN~1940E. Positions of the 768 elliptical coherent beams overlaid on NGC~253 shown in the lower panel.}
    \label{fig_with_beams_and_without_beams}
\end{figure}

\section{TRAPUM Observation} \label{sec:observation}

We observed NGC~253 on 5 October 2022 for a total integration time of 14085.3 seconds. The observation used 44 MeerKAT antennas within the approximately 1~km baseline to create coherent beams that were big enough to cover the entire galaxy, and was carried out at \textit{L}-band, with a central frequency of $f_{\rm c}= 1.284$~GHz, and a bandwidth of $\Delta f = 856$~MHz. The frequency band was divided into 2048 frequency channels and data were sampled every 153~$\upmu$s. Beamforming was performed using the Filterbanking BeamFormer User-Supplied Equipment (FBFUSE; \citealt{Barr_2017}), which created 768 synthesised elliptically-shaped coherent beams for the observation. The shape and orientation of the coherent beams are dependent on the position of the target \citep{2021JAI....1050013C}. We used \mosaic{}\footnote{\url{https://gitlab.mpifr-bonn.mpg.de/wchen/Beamforming/tree/master/mosaic}} \citep{2021JAI....1050013C}, a Python package that simulates one coherent beam with the observational parameters and tiles a specified number of beams at various positions and overlaps them appropriately at a chosen sensitivity level. The tiling pattern is illustrated in the lower panel of \autoref{fig_with_beams_and_without_beams}. The neighbouring beams overlapped at 75 per cent of their peak sensitivity, which ensures that coherent sensitivity is not degraded by more than this amount across the searched area. Data processing was performed on the Accelerated Pulsar Search User-Supplied Equipment (APSUSE; \citealt{Barr_2017}), which is a dedicated pulsar processing computer cluster. We performed a periodicity search with \peasoup{}\footnote{\url{https://github.com/ewanbarr/peasoup}}  (\citealt{2020ascl.soft01014B}, described in \citealt{10.1093/mnras/sty3328}), which is a GPU-based pulsar search algorithm that implements various processing steps such as dedispersion, low-frequency noise removal, and applying a Fast Fourier Transform (FFT). The mitigation of radio frequency interference (RFI) was performed with \pulsarx{}'s\footnote{\url{https://github.com/ypmen/PulsarX}} \filtool{} \citep{2023A&A...679A..20M}. We generated a dedispersion plan with \presto{}'s\footnote{\url{https://github.com/scottransom/presto}} \texttt{DDplan.py} script \citep{2011ascl.soft07017R}, which was then used to efficiently dedisperse the data based on the observational parameters. \peasoup{} uses the \dedisp{}\footnote{\url{https://github.com/ewanbarr/dedisp}} package (part of the \peasoup{} suite; \citealt{2012PhDT.......306L,10.1111/j.1365-2966.2012.20622.x}), to perform dedispersion up to a value of 500~\dmunits. The expected DM along the line of sight for NGC~253 is approximately 19~\dmunits\, and 29~\dmunits\, based on the YMW16 and NE2001 electron density models respectively \citep{Yao_2017,cordes2003ne2001i}.
An acceleration search for pulsars in binaries up to $\pm$ 20 ms$^{-2}$, assuming constant acceleration, was performed. These parameters were chosen in view of computing constraints. An acceleration tolerance of 10 per cent was chosen, implying that the acceleration broadening from one acceleration step to the next cannot exceed 10 per cent of the combined pulse smearing \citep{2012PhDT.......306L}. Furthermore, the candidates were harmonically summed by \peasoup{} to the fourth harmonic. We also applied  default channel and period masks representative of RFI sources that are known to corrupt the MeerKAT L-band. Candidates were then clustered together by \peasoup{} and candidates with periodicities corresponding to known interference were removed. This was followed by a two-stage multibeam candidate filtering performed by \candidatefilter{}\footnote{\url{https://github.com/prajwalvp/candidate_filter}}, with a candidate spectral signal-to-noise (S/N) threshold of 9.5. Furthermore, candidates with similar periodicities were clustered together, and a spatial domain multibeam coincidencing algorithm was applied. This algorithm is used to characterise potential pulsar detections between adjacent beams and distinguish them from RFI sources that are spread over numerous neighbouring beams \citep[see][]{2023MNRAS.524.1291P}.
The remaining candidates were folded at the parameters (DM, spin period, acceleration) obtained from \peasoup{}, using \texttt{psrfold\_fil} from \pulsarx{} \citep{2023A&A...679A..20M} and scored with \textsc{pics}, a Pulsar Image-based Classification System based on Machine Learning that was trained on data from the PALFA survey \citep{Zhu_2014}. \textsc{pics} generates a number between 0 and 1 for each candidate, with 1 representing potential pulsar detection and 0 RFI or noise. We cut the candidates up to a minimum score of 0.1. This left us with 2\,~415 pulsar candidates viewed with \candyjar{}\footnote{\url{https://github.com/vivekvenkris/CandyJar}}. The S/N distribution of these candidates are consistent with the tail of a noise distribution, with candidates outside of this range identified as RFI. \citet{2023MNRAS.524.1291P} contains additional details on the TRAPUM search pipeline.
Furthermore, we also conducted a single pulse search using \transientx{}\footnote{\url{https://github.com/ypmen/TransientX}}, which is a high-performance single pulse search software \citep{2024A&A...683A.183M}. \transientx{} is designed to search successive data blocks, typically a few seconds in duration. Each data block undergoes various processing stages, such as mitigating RFI, dedispersion, matched filtering, clustering, and plotting (see \citealt{2024A&A...683A.183M} for details on the algorithm).
We searched for pulses with a width up to 0.1~s over a dispersion measure range 0--5000~\dmunits\, to accommodate high DMs at which these transients can be detected. A S/N threshold of 8.0 was applied and candidates above this threshold were retained for viewing. We further reduced the number of candidates by removing four short time segments containing RFI. This left us with 2\,314 candidates and the diagnostic plots (showing the single pulse summed over all frequencies, a greyscale of the intensity of the pulse as a function of frequency and a greyscale of the S/N of the pulse vs dispersion measure and time) generated by \transientx{} were viewed manually. All viewed candidates were consistent with noise or RFI.

\begin{figure}
    \centering
    \includegraphics[width =0.49\textwidth]{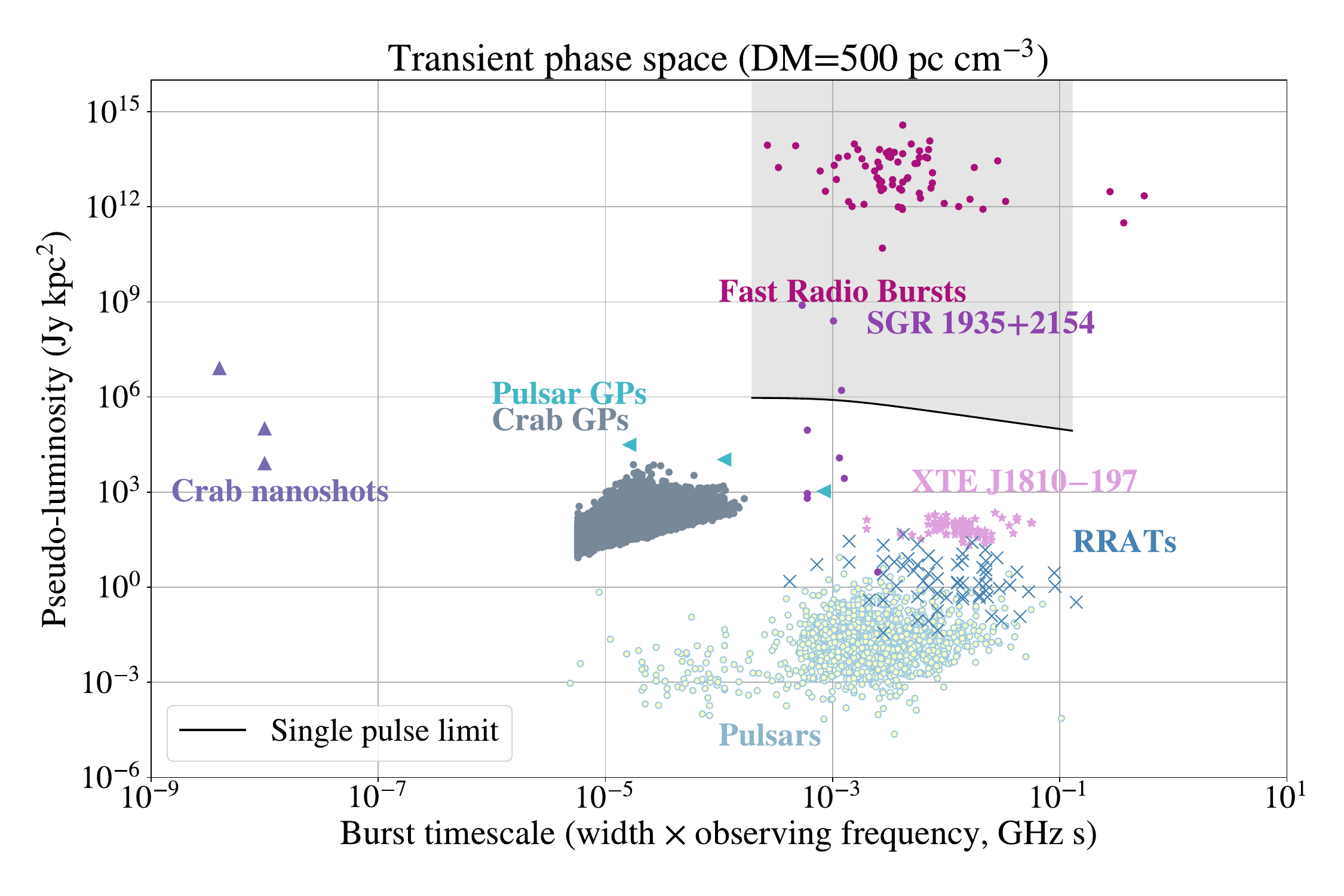}
    \caption{Transient phase space diagram: Illustrated in the shaded region is the pseudo-luminosity search limit for our single pulse search. Noticeably, this observation is sensitive to FRBs, including FRB-like bursts of the same pseudo-luminosity as SGR 1935+2154; however, the detection of ordinary pulsars and giant Crab-like pulses is shown to be improbable for our search due to the large distance of NGC~253.
    Figure and data courtesy of Manisha Caleb.}
    \label{fig:phase-space-diagram}
\end{figure}

\section{Search sensitivity and upper limits} \label{sec:resutls}
Our pulsar search returned no significant candidates during the 4-hour observation. We therefore derived a flux density limit for our pulsar search using a modified radiometer equation from \citep{1985ApJ...294L..25D}:
\begin{equation}
 {S}_{\rm lim} = \beta \frac{{(S/N)} { T}_{\rm sys}}{\epsilon { G}\sqrt{{n_{\rm p}} { t_{\rm int}} \Delta {f}}} \sqrt{\frac{{D}}{1-{D}}},
\end{equation}
where S/N is the spectral S/N threshold of 9.5, $\beta$ is the correction factor due to digitization. Since the loss of telescope sensitivity due to 8-bit digitization is minimal, we have $\beta$ = 1 (\citealt{2001A&A...378..700K}).A conversion efficiency factor $\epsilon = 0.7$ for converting from spectral S/N to folded S/N \citep{2020MNRAS.497.4654M}. The system temperature is the sum of the receiver temperature ($T_{\rm rec}$ = 18~K, \citealt{bailes2020meerkat}) and the sky temperature contribution $T_{\rm sky} \approx $ 4.24~K \citep{2017MNRAS.464.3486Z}\footnote{\url{http://github.com/jeffzhen/gsm2016}} in the direction of NGC~253. The telescope gain is G = 1.925~K Jy$^{-1}$ for the core array in coherent mode (G = 0.34~${\rm K Jy^{-1}}$ for incoherent mode with 61 antennas). Lastly, the number of polarisations is $n_{\rm p} = 2$, the bandwidth $\Delta f = 856$~MHz, and the integration time $t_{\rm int}$ = 14085.30~s. 
We assume a pulsar duty cycle of $D=2.5\%$, which is the median intrinsic pulsar duty cycle excluding millisecond pulsars and pulsars in globular clusters, rotating radio transients, magnetars and binary pulsars, taken from the ATNF catalogue \citep{manchester2005australia}, and a spectral index of $-1.6$ \citep{2018MNRAS.473.4436J}. Thus, the minimum detectable flux density for our periodicity search is $S_{\rm lim,\rm 1400MHz}$= 4.4~$\upmu$Jy at the centre of the incoherent beam and of a coherent beam for a pulsar with a period of 100~ms and above at a DM = 250~\dmunits. At lower spin periods, factors such as dispersion smearing, sampling time, and scattering influence the detectability of such short period pulsars. The corresponding pseudo-luminosity limit, which is calculated using
\begin{equation}
L_{1400} = {S}_{1400}{d}^2,
\end{equation}
is about 54~Jy~kpc$^2$ for a distance of $d=3.5$~Mpc.

For our fast transient search, none of the remaining candidates yielded any promising detections for FRBs or FRB-like pulses. Therefore, we also derived the sensitivity limit of our transient search by using the radiometer equation 
\begin{equation}
    {S}_{\rm pulse\, peak} = \frac{{(S/N)} {S}_{\rm sys}}{\sqrt{{n}_{\rm p} \Delta {f} W }},  
\end{equation}
(e.g. \citealt{2003ApJ...596..982M}). Here $W$ is the observed width of the pulse. We assume an intrinsic pulse width of 1~ms at a DM of 500~\dmunits. We also take into consideration the dispersive smearing across individual channels and the smearing due to sampling time. S$_{\rm sys}$ is the equivalent system flux density ($T_{\rm sys}/G$). We calculated a single pulse flux density limit of $S_{\rm peak, 1,284 \rm GHz}\simeq$62~mJy. \autoref{fig:phase-space-diagram} shows which astronomical objects our single pulse search is sensitive to. We used a DM = 500~pc cm$^{-3}$, a coherent mode gain, and pulse widths from $t_{\rm samp}$ to 0.1~s for the search sensitivity in \autoref{fig:phase-space-diagram}. It can be seen from \autoref{fig:phase-space-diagram} that our search was sensitive to FRBs and FRB-like bursts such as some of those from SGR~1935+2154.  
 \begin{figure}
     \centering
     \includegraphics[width=0.5\textwidth]{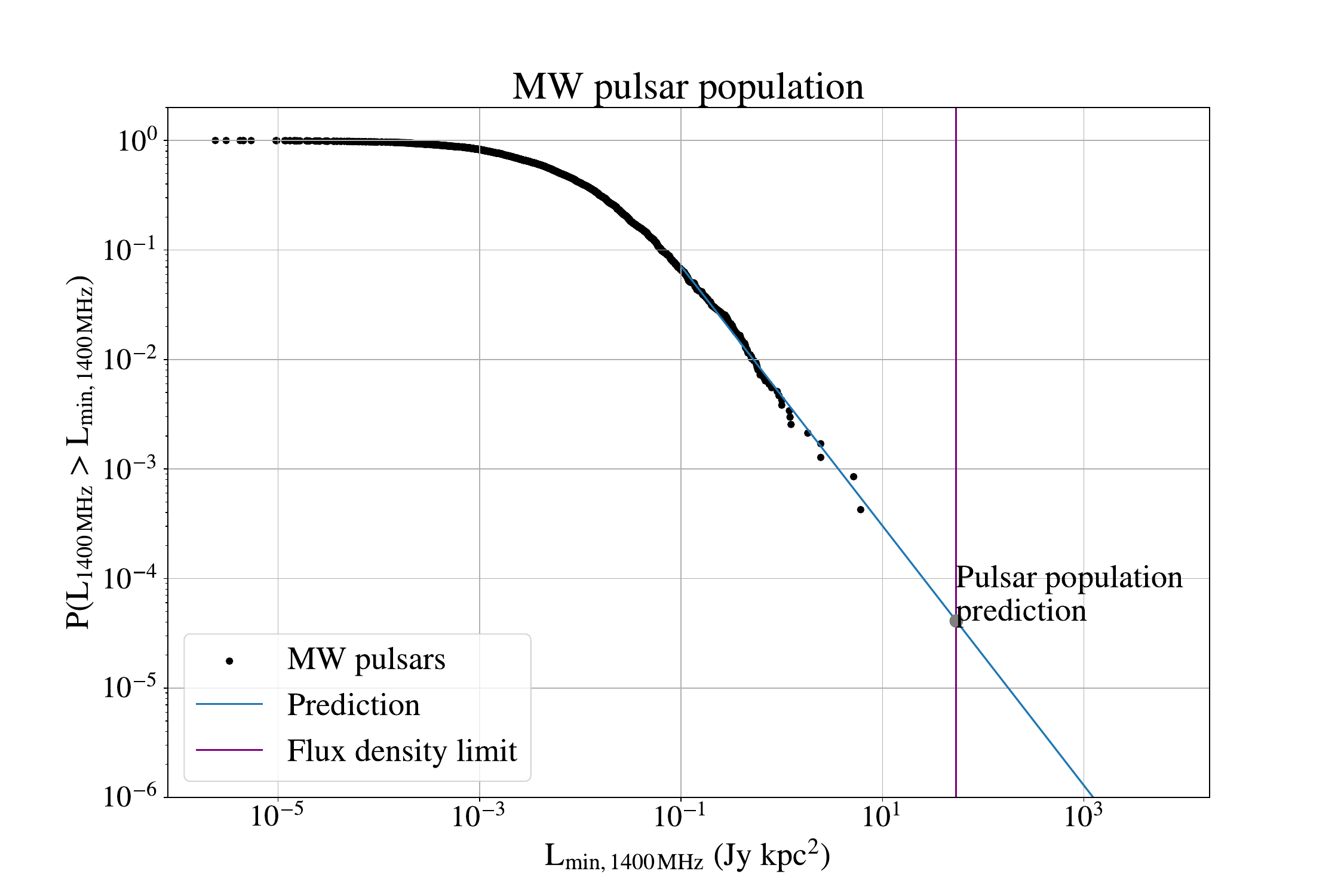}
     \caption{Pseudo-luminosity distribution of the MW pulsar population compared to our survey search limits. Figure produced with ATNF catalogue version 2.0.0.}
     \label{fig:psr-pop-diagram}
\end{figure}

\section{Discussion \& Conclusion} \label{sec:discussion_and_conclusion}
 
NGC~253 is an intriguing target for extragalactic pulsar exploration due to its stellar composition, current star formation, host SNRs and associated MGFs seen towards this galaxy. The presence of the optically identified SN~1940E is especially interesting for a young pulsar search, since the lifetime of SNRs is considerably shorter than that of pulsars (e.g., \citealt{lyne2012pulsar}). Thus, a population of young pulsars is expected in this galaxy. Such pulsars are more probable to be detected at vast extragalactic distances, since these objects may emit giant single pulses \citep{2003ApJ...590L..95J}, the flux density of which is orders of magnitude higher than that of ordinary pulsars. We therefore searched for pulsars, magnetars, and FRBs in NGC~253 with MeerKAT. However, \autoref{fig:phase-space-diagram} shows that our single-pulse search was more likely to detect FRBs or equivalent bright-emitting objects (such as SGRs) and that we were not sensitive to the detection of MW-type giant pulses from young pulsars. 

Surveys such as \textit{The Parkes Southern Pulsar Survey} \citep{1996MNRAS.279.1235M}  and \textit{The High Time Resolution Universe Pulsar Survey} \citep[HTRU; ][]{2010MNRAS.409..619K}  included pointings towards NGC~253, the integration time of which were $157.3$~s and  $270$~s respectively. We searched the entire NGC~253 galaxy, which increased our chances of finding a pulsar as compared to searching only portions of this galaxy. Our integration time is about 52 times longer than that of \textit{HTRU} and our gain is also a factor of 3 times more. Thus, our search with MeerKAT was significantly more sensitive than previous surveys with the Parkes Telescope.  

Our periodicity search towards NGC~253 returned no significant candidates. We derived a flux density limit for periodic pulsar signals of $S_{\rm lim,\rm 1400MHz}$ = 4.4~$\upmu$Jy, which corresponds to a pseudo-luminosity limit of $L_{\rm lim,\rm 1400MHz}=$ 54 Jy kpc$^2$. In \autoref{fig:psr-pop-diagram}, we show the cumulative probability distribution of the known MW pulsars from the ATNF catalogue as a function of pseudo-luminosity, with our sensitivity limit indicated by a purple vertical line. The brightest MW pulsar pseudo-luminosity at 1400~MHz from the ATNF catalogue is about 6 Jy kpc$^2$ (J1644$-$4559), which is one order of magnitude lower than this limit. Since the pulsar population in the MW spans about 6 orders of magnitude in pseudo-luminosity, and the maximum pseudo-luminosity is not known, we may be able to probe the upper range of this function. If this luminosity function is taken to be somewhat universal (since NGC~253 is experiencing a period of rapid star formation and has a mass difference of about 50 per cent compared to the MW), it is therefore not unreasonable to search for pulsars in NGC~253. Fitting a power law above $\sim0.1~{\rm Jy\,kpc}^2$ ($\sim0.25~{\rm Jy\,kpc}^2$,$\sim0.6~{\rm Jy\,kpc}^2$) and extrapolating, given the shape of the tail of the distribution at this point, we estimate that the chance of finding a pulsar as bright as our limit is approximately $4.10\times10^{-5}$ ($2.51\times10^{-5}$, $4.25\times10^{-5}$).  Thus, in order to detect such a pulsar in NGC~253, this galaxy should have a pulsar population of at least $2.4\times10^{4}$ ($3.9\times10^{4}$, $2.3\times10^{4}$). This is not an unreasonable number, and is a good fraction of the estimated total population of detectable MW pulsars of 122 000 (including millisecond and normal pulsars; \citealt{2014ApJ...787..137S}). This implies that approximately 5 MW pulsars as bright as our limit should exist (slightly more if one integrates up to a certain maximum pseudo-luminosity to include all those exceeding the current (vertical line in \autoref{fig:psr-pop-diagram}) sensitivity limit). However, no such pulsars have been detected to date. This might be due to unfavourable beaming, excessive DM smearing, or obstruction by other sources. Alternatively, the luminosity distribution might cut off earlier (departing from the power-law tail), implying a strict upper limit on the maximum pseudo-luminosity of $L_{\rm cut,1400MHz}\lesssim L_{\rm lim,1400MHz}$ for both galaxies, and in this case we should not expect to see such a bright pulsar in NGC~253. Otherwise, the luminosity distributions may be different between the galaxies, reflecting distinct star-formation histories. 
MW pulsars have typical conversion efficiencies ($\eta$) of radio emission derived from the pulsar rotational spin-down luminosity $\dot{E}_{\rm rot}$ of $10^{-8}-10^{-5}$ (e.g. \citealt{Szary_2014}). Given our periodic-signal sensitivity limit, a Crab-like pulsar with $\dot{E}_{\rm rot}\sim5\times10^{38}$~erg\,s$^{-1}$ should have been detectable if $\eta\gtrsim10^{-8}$. Conversely, for a typical value of $\eta\sim10^{-5}$, a pulsar with $\dot{E}_{\rm rot}\sim5\times10^{35}$~erg\,s$^{-1}$ should have been detectable.

Moving to single-pulse searches, as seen in \autoref{fig:phase-space-diagram}, we were sensitive to FRB and FRB-like pulse detections. We derived a single-pulse flux density limit of approximately 62~mJy for our FRB search in the direction of NGC~253. This limit is orders of magnitude lower than the flux density of previously detected FRBs. Since two MGFs are associated with NGC~253 \citep{2024A&A...687A.173T,2021Natur.589..211S}, we would expect to be able to observe bursts similar to those of SGR~J1935+2154 in the MW, should they have occurred. However, our non-detection can potentially be due to the epoch of observation, since these events are unpredictable and sporadic. It can also be due to unfavourable orientations. Transient radio emission, lasting for years have also been seen from magnetars, but none has been associated with MGFs, since these events are rare and not much radio follow up has been possible \citep{2006Natur.442..892C,2019MNRAS.488.5251L}
 
Future observations covering a longer integration time may uncover pulsars and FRBs in this and other nearby galaxies.

%%%%%%%%%%%%%%%%%%%%%%%%%%%%%%%%%%%%%%%%%%%%%%%%%%%%%%%
\section*{Acknowledgements}
We thank the anonymous referee for their constructive feedback that helped improve the quality of the paper.
The MeerKAT telescope is operated by the South African Radio
Astronomy Observatory, which is a facility of the National Research
Foundation (NRF), an agency of the Department of Science and Innovation (DSI). SARAO acknowledges the ongoing advice and calibration of
GPS systems by the National Metrology Institute of South Africa
(NMISA) and the time space reference systems department of the 
Paris Observatory.
TRAPUM observations used the FBFUSE and APSUSE computing clusters for data acquisition, storage and analysis. These clusters were designed, funded and installed by the Max-Planck-Institut für Radioastronomie and the Max-Planck Gesellschaft. 
This work is based on research supported in part by the NRF. C.V.\ acknowledges that opinions, findings and conclusions or recommendations expressed in any publication generated by the NRF-supported research are those of the author(s), and that the NRF accepts no liability whatsoever in this regard.
H.H.\ acknowledges research support in part by the NRF of South Africa (Grant number 142 859), and support through the National Astrophysics and Space Science Program (NASSP), funded by NRF. 
R.P.B.\ acknowledges support from the European Research Council (ERC) under the European Union's Horizon 2020 research and innovation program (grant agreement No. 715051; Spiders)
This research has made use of hips2fits,\footnote{https://alasky.cds.unistra.fr/hips-image-services/hips2fits} a service provided by CDS, the SIMBAD database,
operated at CDS, Strasbourg, France \citep{2000A&AS..143....9W}, NASA's \href{https://ui.adsabs.harvard.edu}{Astrophysics Data System} Bibliographic Services, and the ATNF pulsar catalogue version 2.0.0 and 2.3.0.
\autoref{fig_with_beams_and_without_beams} was made using the APLpy v2.1.0 \citep{aplpy2012,aplpy2019}, which is an open-source astronomy plotting Python package. 

%%%%%%%%%%%%%%%%%%%%%%%%%%%%%%%%%%%%%%%%%%%%%%%%%%
\section*{Data Availability}

The data underlying this article will be shared upon reasonable request to the TRAPUM collaboration.



\bibliographystyle{mnras}
\bibliography{ngc253_bib_1} % if your bibtex file is called example.bib

%%%%%%%%%%%%%%%%%%%%%%%%%%%%%%%%%%%%%%%%%%%%%%%%%%

% %%%%%%%%%%%%%%%%% APPENDICES %%%%%%%%%%%%%%%%%%%
%%%%%%%%%%%%%%%%%%%%%%%%%%%%%%%%%%%%%%%%%%%%%%%%%%

% Don't change these lines
\bsp	% typesetting comment
\label{lastpage}
\end{document}